\documentclass[11pt]{article}
\usepackage{epstopdf}
\usepackage[a4paper,hmarginratio=1:1,vmarginratio=2:3,totalwidth=15.3cm,totalheight=23cm]{geometry}
\usepackage{bm,epstopdf,epsfig,amsmath,amssymb,amsfonts,colordvi,latexsym,
comment,cancel,verbatim,%extarrows,
slashed}
\usepackage{epsfig,bbm}
\usepackage{graphicx,graphics}
\usepackage[font=md,captionskip=8pt]{subfig}
\usepackage[usenames,dvipsnames]{color}
\usepackage[noadjust]{cite}
\usepackage{xcolor}  
\usepackage{setspace}
\newcommand{\h}{\sigma}

\newcommand{\sg}{\sqrt{g}}

\newcommand{\w}{\omega}
\usepackage[utf8]{inputenc}
\allowdisplaybreaks
\newcommand{\cL}{{\cal L}}

\newcommand{\cO}{{\cal O}}   

\newcommand{\cR}{{\tilde R}}

\newcommand{\cV}{{\mathcal V}}

\newcommand{\ra}{\rightarrow}

\newcommand{\be}{\begin{equation}}
\newcommand{\ee}{\end{equation}}
\newcommand{\bea}{\begin{eqnarray}}
\newcommand{\eea}{\end{eqnarray}}

\newcommand{\baa}{\begin{array}}
\newcommand{\eaa}{\end{array}}

\long\def\symbolfootnote[#1]#2{\begingroup
\def\thefootnote{\fnsymbol{footnote}}\footnote[#1]{#2}\endgroup}
\begin{document} 
\begin{flushright}
%{ \today}
\end{flushright}
\bigskip\medskip
\thispagestyle{empty}
\vspace{2.2cm}
\begin{center}

{\Large \bf  Stueckelberg breaking of  Weyl conformal geometry

\bigskip
and applications to gravity}

\vspace{1.cm}

 {\bf D. M. Ghilencea} \symbolfootnote[1]{E-mail: dumitru.ghilencea@cern.ch}

\bigskip
{\small Department of Theoretical Physics, National Institute of Physics
 
and Nuclear Engineering, Bucharest\, 077125, Romania}
\end{center}

\bigskip
\begin{abstract}
\begin{spacing}{1.07}
\noindent
Weyl conformal geometry  may play a role in early  cosmology where effective 
theory at  short distances  becomes conformal. Weyl conformal geometry also 
has a built-in geometric Stueckelberg  mechanism: it is  broken spontaneously 
to  Riemannian geometry after a particular Weyl gauge transformation (of ``gauge fixing'') 
while Stueckelberg mechanism re-arranges the  degrees of freedom,  conserving their number ($n_{df}$).  
The Weyl gauge field ($\w_\mu$) of local scale transformations  acquires a  mass 
 after absorbing a  compensator (dilaton), decouples, and  Weyl  connection becomes Riemannian.
Mass generation has thus a dynamic  origin, corresponding to a
 transition from  Weyl to  Riemannian  geometry.
We show that
a  ``gauge fixing'' symmetry transformation of the   original  Weyl's  quadratic 
gravity action in its Weyl geometry formulation,
 immediately gives the Einstein-Proca action for the Weyl gauge  field and a positive 
cosmological constant, plus matter action (if  present). 
As a result, the Planck scale is an {\it emergent} scale, where Weyl gauge symmetry is 
spontaneously broken and Einstein action is the broken phase of Weyl  action.
This is in contrast to local scale invariant  models (no gauging) where a
negative kinetic term  (ghost dilaton) remains present 
 and $n_{df}$ is not conserved when this symmetry is broken.
The mass of $\w_\mu$, setting  the non-metricity scale, can  
be much  smaller than $M_\text{Planck}$, for  ultraweak values of the coupling ($q$),   
with  implications  for phenomenology.  If matter is present, a {\it  positive} 
contribution to the Planck scale from a  scalar field ($\phi_1$) vev
induces  a {\it negative}  (mass)$^2$  term for $\phi_1$ and spontaneous breaking 
of the symmetry under which it is charged. These results are 
immediate when using a Weyl geometry formulation of an action instead of its Riemannian
picture. Briefly, Weyl gauge symmetry is  physically relevant and its
 role in high scale physics  should be reconsidered.
\end{spacing}
 \end{abstract}

\newpage

\section{Weyl gauge  transformations and Stueckelberg mechanism}

 In 1918 Weyl introduced  his 
vector-tensor theory of quadratic gravity  \cite{Weyl,Weyl2,Weyl3} 
built  on what is now known as Weyl conformal geometry.
Weyl's idea was that the action should be invariant 
under a most general symmetry: a Weyl  scaling gauge  symmetry  \cite{Scholz}.
Weyl also thought of identifying this  gauge field ($\omega_\mu$) with electromagnetism,
which inevitably  failed since electromagnetic gauge transformations 
are  ``internal''   symmetry (not spacetime geometry) transformations.
 Weyl quadratic gravity was  disregarded after Einstein's early
 criticism \cite{Weyl} that
the spacing of atomic spectral lines  changes in such theory, in contrast with
experience. This happens because in Weyl geometry
a vector parallel transported around a 
 curve changes not only the direction (as in Riemannian geometry) but also its
length. Then clock's rates and rod's lengths  depend on their path history.
This is caused by the  {\it massless} Weyl gauge field $\omega_\mu$
responsible for the non-metric connection  of Weyl  geometry, 
$\tilde\nabla_\mu g_{\alpha\beta}\!=\!-\omega_\mu\, g_{\alpha\beta}$. This is in contrast to
Riemannian case (of $\w_\mu\!=\!0$) and Einstein gravity %(of $\w_\mu\!=\!0$), %% giving
 where $\nabla_\mu \,g_{\alpha\beta}\!=\!0$ with $\nabla_\mu$  the Levi-Civita connection.
Eventually (gauged) local scale transformations were abandoned and  replaced by  phase 
transformations \cite{London} setting  the foundation  of modern gauge theories.

Dirac revived Weyl gravity by introducing  a different version of it \cite{D}
{\it linear}  in Weyl scalar curvature ($\cR$) of the form $\phi^2 \cR$
with an {\it additional}  matter scalar $\phi$  \cite{Oh,Sm2,Tang,H2,M,N2,C2,S2,Q2,ghilen,J2,T2}.
 This term recovers  Einstein gravity, the Weyl field becomes massive
(mass $\sim\! q M_\text{Planck}$) and decouples ($\w_\mu\!=\!0$); as a result,
Weyl connection  becomes  Riemannian and Einstein's criticism  is avoided.

Recently it was shown \cite{G} that even the original Weyl {\it quadratic} gravity 
{\it without} matter \cite{Weyl,Weyl2,Weyl3} avoids  Einstein's criticism  since %the field 
$\w_\mu$ again becomes massive and decouples.
Here  we explore further the consequences of  this work.
The result in \cite{G} underlines the less known fact that
 theories based on Weyl  geometry % such as the original Weyl quadratic gravity
have  a built-in {\it geometric} Stueckelberg mass  mechanism \cite{S,Ruegg,RP1}. 
To see this more easily and unlike in \cite{G}, here  we use the Weyl geometry formulation of an action
instead of the Riemannian picture. This simplifies our calculations.

The main new results of this work show that: 

\noindent
{\bf a)} in the Weyl formulation of an action
 a simple Weyl gauge symmetry transformation (of  ``gauge fixing'')  applied to it
gives an action directly in the Riemannian geometry 
with a  Stueckelberg breaking of the Weyl gauge  symmetry.
For example, the original  Weyl quadratic gravity action is ``gauge transformed'' 
into  Einstein-Proca action for $\w_\mu$, a cosmological constant
plus matter action (if  present). So
Einstein action is just a {\it spontaneously broken phase} of  Weyl quadratic gravity action
(in the absence of matter).

\noindent
{\bf b)} We stress that only gauge transformations are used in\,step {\bf a)}
but  no fields re-definitions.

\noindent
{\bf c)} Note that no ghost is generated and  the number $n_{df}$ 
of degrees of freedom (other than graviton)
is conserved: the real dilaton (spin 0  mode in the $\tilde R^2$ term)
 is absorbed by  $\w_\mu$ which becomes massive; so $n_{df}\!=\!3$ is unchanged,
 as expected for a spontaneous  breaking.
This is different from ``gauge fixing'' in conformal models (e.g.\cite{TH2}) where 
Stueckelberg mechanism is not available (since there is no $\w_\mu$) so $n_{df}$ 
is not conserved and a ghost dilaton is present;

\noindent
 {\bf d)} Planck scale is an {\it emergent scale} where Weyl gauge symmetry is 
spontaneously broken.

The non-metricity scale is set by the Weyl ``photon'' mass ($\sim q\, M_\text{Planck}$), 
naively expected to  be  large. Interestingly,  small values of this  mass are  allowed  
(demanding ultraweak values of its coupling $q$)  because the lower bound on 
non-metricity scale is $\cO(\text{TeV})$ \cite{La}. Then  the Weyl field could
even be a (TeV) dark matter candidate \cite{Ta}.  The  phenomenology 
of Standard Model (SM)  endowed with  Weyl gauge symmetry   \cite{M,ghilen} deserves careful study.

\subsection{Weyl gauge transformations}

 Consider  a Weyl scaling gauge transformation  $\Omega(x)$ of  the metric 
$g_{\mu\nu}$    and of  scalar  field $\phi$
\bea
\label{ct}
 \hat g_{\mu\nu} =  \Omega\, g_{\mu\nu},\qquad
\hat\phi  =  \frac{1}{\sqrt \Omega}\,\phi, 
\qquad
\hat\w_\mu =\w_\mu-\partial_\mu\ln\Omega.
\eea
Here $\omega_\mu$ is the Weyl gauge field; 
% $q$ is the coupling to $\phi$;
we also have $\sqrt{\hat g}\!=\! \Omega^2\! \sg$,\,
 $g\!\equiv\!\vert\det g_{\mu\nu}\vert$ and metric $(+,-,-,-)$ and conventions as in \cite{R}.
The Weyl-covariant derivative of $\phi$ is 
\bea
\label{ST}
\tilde D_\mu \phi&=&(\partial_\mu-1/2\,\,\w_\mu)\,\phi
\\[-3pt]
& =&
(-1/2)\,\phi\, \big[\,\w_\mu- \partial_\mu\ln \phi^2\,\big].
\label{ST2}
\eea
 $\Omega(x)$ is real, there is no complex factor ``i'' in 
(\ref{ct}) or in $\tilde D_\mu\phi$.
 The gauge symmetry  is a  dilatation group which is isomorphic to $R^+$.
$\tilde D_\mu\phi$  transforms under (\ref{ct}) like a  scalar field 
$\hat{\tilde{D}}_\mu\hat\phi=(1/\sqrt{\Omega})\,\tilde D_\mu\phi$.
Given  (\ref{ct}),  $\w_\mu$ has  geometric origin while eq.(\ref{ST2})  has 
an obvious resemblance to  the Stueckelberg mechanism, see later.

In Weyl geometry   $(\tilde\nabla_\mu+\w_\mu)\, g_{\alpha\beta}=0$, with $\tilde\nabla_\mu$
defined by the Weyl connection coefficients denoted $\tilde \Gamma_{\mu\nu}^\rho$. 
This differs from  Riemannian geometry where $\nabla_\mu g_{\alpha\beta}=0$ with
$\nabla_\mu$ defined by the Levi-Civita connection\,\, $\Gamma_{\mu\nu}^\rho = 
(1/2)\, g^{\rho\beta}\,(\partial_\nu g_{\beta\mu}+\partial_\mu g_{\beta\nu}-\partial_\beta g_{\mu\nu})$.
$\tilde \Gamma_{\mu\nu}^\rho$  can be found from $\Gamma_{\mu\nu}^\rho$ by  replacing
  $\partial_\mu\ra \partial_\mu +\w_\mu$  and
giving
$\tilde\Gamma_{\mu\nu}^\rho=
\Gamma_{\mu\nu}^\rho+(1/2)\,\big[\delta_\mu^\rho\,\w_\nu+\delta_\nu^\rho\,\w_\mu-g_{\mu\nu}\,\w^\rho\big]$.

 $\tilde \Gamma_{\mu\nu}^\rho$ are symmetric ($\tilde \Gamma_{\mu\nu}^\rho=\tilde \Gamma_{\nu\mu}^\rho$)
(no torsion) and  
are invariant under (\ref{ct}) since their variation induced by the metric is compensated by that of 
$\w_\mu$. The Riemann and Ricci  tensors  in Weyl geometry are defined as 
in  Riemannian geometry but in terms of new $\tilde \Gamma_{\mu\nu}^\rho$, and are also 
invariant under (\ref{ct})\footnote{These are $ \cR^\lambda_{\,\mu\nu\sigma}=
 \partial_\nu \tilde\Gamma^\lambda_{\mu\sigma}
 -\partial_\sigma\tilde\Gamma^\lambda_{\mu\nu}
 + \tilde\Gamma^\lambda_{\nu\rho}\,\tilde\Gamma^\rho_{\mu\sigma}
 -\tilde\Gamma^\lambda_{\sigma\rho}\,\tilde\Gamma^\rho_{\mu\nu},
$ and $\cR_{\mu\sigma}=\cR^\lambda_{\,\mu\lambda\sigma}$. Also we have:
$\cR=g^{\mu\sigma}\,\cR_{\mu\sigma}$. 
}. 
One can then show that the Weyl scalar curvature ($\cR$) 
\medskip
\bea\label{tildeR}
\cR =
R-3 \,   D_\mu \w^\mu -\frac32 \, \w^\mu \w_\mu,
\eea

 \medskip\noindent
where $R$ is the  Riemannian scalar curvature 
and $D_\mu\w^\mu$ is defined by Levi-Civita connection.

Using the  curvature tensors and scalar of Weyl geometry
has an advantage:  unlike in Riemannian case,  $\cR$ 
transforms  covariantly  under (\ref{ct})  similar to $g^{\mu\nu}$ 
entering its definition: 
\bea\label{RR}
\hat{\cR}= \frac{1}{\Omega}\,\cR.
\eea
This simplifies our calculations and helps build Weyl gauge  invariant 
individual operators using the Weyl formulation of an action.
 Then this symmetry and internal gauge symmetries
are on an equal footing in the action.

The criticisms of Weyl gravity based on Weyl geometry (such as the change of a
vector length under parallel displacement %around  a closed loop 
or of atomic spectral lines spacing) are avoided if $\w_\mu\!=\!0$
because from above $\tilde \Gamma_{\mu\nu}^\rho\!=\!\Gamma_{\mu\nu}^\rho$, $\tilde R=R$,
 Weyl connection becomes Levi-Civita, the
geometry is Riemannian, and  these criticisms do not apply.
This happens  if we do not introduce $\w_\mu$ in (\ref{ct}) i.e. 
we go back to Riemannian geometry gravity (e.g. Brans-Dicke).
Alternatively,  $\w_\mu\!=\!0$  after this field acquires  a large mass 
and decouples.  This is the idea we study below.

\subsection{Weyl gauge transformation and Proca action} \label{WP}

 Consider $L$  of a real
scalar field $\phi$ with coupling $q$ to a Weyl gauge field $\w_\mu$, invariant under~(\ref{ct})
%\smallskip
\bea\label{ty}
L=\sqrt g \,\Big[ -\frac{1}{4 q^2} F_{\mu\nu}^2 + \frac12 \,(\tilde D_\mu\phi)^2 % -\lambda\,\phi^4
\Big].
\eea

\medskip\noindent
To simplify  notation, we do not show
appropriate indices  contractions  which are implicit, e.g.:
 $F_{\mu\nu}^2=g^{\mu\nu} g^{\rho\sigma} F_{\mu\rho} F_{\nu\sigma}$ and 
$(\tilde D_\mu\phi)^2=g^{\mu\nu} \tilde D_\mu\phi\tilde D_\nu\phi$, etc.
Since there is no torsion, the field strength $F_{\mu\nu}$
 does not feel the connection. From
 $F_{\mu\nu}=\tilde D_\mu\w_\nu-\tilde D_\nu\w_\mu$ with 
$\tilde D_\mu\w_\nu=\partial_\mu\w_\nu-\tilde\Gamma_{\mu\nu}^\rho \w_\rho$
% =(\partial_\mu\w_\nu-\tilde\Gamma_{\mu\nu}^\rho \w_\rho)
% -(\partial_\mu\w_\nu-\tilde\Gamma_{\mu\nu}^\rho \w_\rho)
then  $F_{\mu\nu}=\partial_\mu \w_\nu-\partial_\nu\w_\mu$
which coincides with its Riemannian expression  and is invariant under (\ref{ct}).
A gauge transformation (\ref{ct}) with $\Omega=\phi^2/M^2$  gives
\medskip
\bea\label{a}
L=\sqrt {\hat g}\,\Big[-\frac{1}{4 q^2}\,\hat F_{\mu\nu}^{2}
+\frac{1}{8}\,M^2 \hat\w_\mu \hat\w^{\mu} 
\Big].
\eea

\medskip\noindent
where $M$ is an arbitrary scale and all indices contractions are  made with 
new metric ($\hat g_{\mu\nu}$).

The Weyl ``photon'' has become massive and no trace of $\phi$ is left,
see \cite{RP1} for a more general discussion.
This is a geometric version of  Stueckelberg mechanism \cite{S,Ruegg}
which is naturally  built-in 
 Weyl conformal geometry due to the definition of Weyl-covariant derivative $\tilde D_\mu$.
%see \cite{RP1} for a general discussion.
The presence of $\sqrt g$ is essential as it  ensures invariance of $L$.  
The (charged) scalar and Weyl kinetic term %in Weyl formulation
 are gauge 
transformed into an equivalent   Proca   action with spontaneous breaking of Weyl 
gauge symmetry. If we do the inverse gauge transformation, Proca action of a massive theory 
can be written in  a Weyl gauge invariant way as a sum of  kinetic terms.

The gauge transformation we did is  essentially ``gauge fixing'' $\phi\!=\!M$ (constant)
but what is  most important here 
is the conservation of the number of dynamical degrees of  freedom, $n_{df}$ ($n_{df}\!=\!3$):
initially we had a massless scalar and a massless vector field and finally a massive vector field. 
$q M$ is regarded as the scale where Weyl gauge symmetry is broken. 

If  in (\ref{ty}) there are more ($n$) scalar fields kinetic terms, consider
a  gauge transformation 
$\Omega=\rho^2/M^2$,  with $\rho$ the radial direction 
$\rho^2\!=\!\sum_j\phi_j^2$,  absorbed by the only vector field present
$\hat\w_\mu\!=\!\w_\mu- \partial_\mu\ln\rho^2$ under (\ref{ct}). One recovers (\ref{a}) with 
$n-1$ additional   kinetic terms for the angular variables fields.
To conclude, the Weyl boson  is  massive and can  decouple.

\subsection{Weyl linear gravity as Einstein-Proca action}

Consider a linear version of Weyl gravity \cite{D}
coupled to a  scalar $\phi_1$, invariant under (\ref{ct})
\medskip
\bea\label{L1}
\cL=
 \sg\,\Big[
- \frac{\xi_1}{12} \,\phi_1^2\,\cR
+\frac12 \,g^{\mu\nu}\,\tilde D_\mu\phi_1 \tilde D_\nu \phi_1 - \frac{\lambda_1}{4!}\phi_1^4
-\frac{1}{4 q^2} F_{\mu\nu}^2\Big],
\eea

\medskip\noindent
where $\cR$ is the scalar curvature in Weyl geometry, eq.(\ref{tildeR}), and
 $\tilde D_\mu\phi_1\!=\!(\partial_\mu-1/2\,\,\w_\mu)\,\phi_1$.

After a Weyl gauge transformation (\ref{ct}), (\ref{RR}),
 with $\Omega=\xi_1\phi_1^2/(6 M^2)$, then 
using eq.(\ref{tildeR})
\medskip
\bea\label{L2}
\cL=
\sqrt{\hat g} \, \Big[
-\frac{1}{2} M^2\,\hat R 
-\frac{1}{4 q^2}  \hat F_{\mu\nu}^{2}
+\frac34\, M^2\, \big(1+1/\xi_1\big)\, \hat\w_\mu\,\hat\w^{\mu} 
-\frac{3\lambda_1  M^4}{2\,\xi_1^2}\Big],
\eea

\medskip\noindent
up to a total derivative term. Here
 Riemannian scalar curvature $\hat R$ and indices contractions are computed with new  $\hat g_{\mu\nu}$,
 as indicated by the presence of $\sqrt{\hat g}$.

The gauge transformation  considered sets $\hat\phi_1$ to a constant  
($6 M^2/\xi_1$), and the Einstein  frame results from ``gauge fixing''  Weyl gauge symmetry.
The Stueckelberg mechanism ensures the number of dynamical degrees of freedom $n_{df}$ is 
conserved when going from (\ref{L1}) to (\ref{L2}) as expected for  spontaneous 
breaking (and which does not require  a potential for  $\phi_1$). Here $\phi_1$ was ``eaten'' 
by the Weyl gauge field which is now massive. What survives of the scalar kinetic 
term is the $\xi$-dependent mass term for $\hat\w_\mu$, but there is  an 
additional  mass correction to $\hat\w_\mu$ beyond (\ref{a}), due to the 
$\tilde R$-dependent term.

This situation is in contrast with  the (ungauged) local
conformal symmetry case  
 recovered from (\ref{L1})  for $\w_\mu=0$;
then there is no gauge field to ``absorb'' the scalar ``compensator''
and  the action would be invariant under the  first two transformations in (\ref{ct}) 
only if $\xi_1=-1$.

To conclude,  Weyl photon again  became  massive by ``absorbing'' a ``compensator'' field
$\phi_1$. But what happens in  Weyl quadratic gravity with no  matter fields present?

\subsection{Weyl quadratic gravity as Einstein-Proca action}\label{QG}

The original  action of Weyl (quadratic) gravity  without  matter \cite{Weyl2}
invariant under (\ref{ct}) is
\medskip
\bea\label{Rsq}
L_1=\sqrt g \,\Big[\,\frac{\xi_0}{4!}\,\cR^2 -\frac{1}{4 q^2} F_{\mu\nu}^2\,\Big],\qquad \xi_0>0.
\eea

\medskip\noindent
Each term is Weyl gauge invariant ($\cR$ transforms covariantly, eq.(\ref{RR})).
We  can replace $\cR^2\ra -2\phi_0^2\,\cR -\phi_0^4$, since integrating the
auxiliary field $\phi_0$ via its equation of motion, of solution $\phi_0^2=-\cR$,
 recovers the  $\cR^2$ term in the action;
so  $\phi_0$ transforms like any scalar field and $\ln\phi_0$ is
 the Goldstone of the scale symmetry (\ref{ct}), $\ln\phi_0^2\ra\ln\phi_0^2-\Omega$. Then
%\medskip
\bea\label{Rsq2}
L_1=\sqrt g \,\Big[\,\frac{\xi_0}{4!}\,\big(-2\phi_0^2\, \cR 
-\phi_0^4\big) - \frac{1}{4 q^2} F_{\mu\nu}^2\,\Big].
\eea

\medskip\noindent
Using  gauge transformation (\ref{ct}), (\ref{RR})  with  $\Omega=\xi_0\,\phi_0^2/(6 M^2)$,
{\it then} using relation (\ref{tildeR}) we find 
\medskip
\bea\label{EP}
L_1=\sqrt{\hat g}\,
\Big\{-\frac{1}{2}\,M^2\,\hat R
-\frac{3\,M^4}{2\xi_0}
+\frac{3}{4}\,M^2
\hat\w_\mu\,\hat\w^{\mu}
-\frac{1}{4 q^2}\,\hat F_{\mu\nu}^{2}
\Big\}.
\eea

\medskip\noindent
which is in  the Einstein frame. Here we chose $M=M_\text{Planck}$;
($\hat R$ is the Riemannian scalar curvature evaluated from new metric $\hat g_{\mu\nu}$
also used for  index contractions).

These simple steps show an interesting result:  
a  Weyl ``gauge fixing'' symmetry transformation (not fields redefinition)
applied to the original  Weyl quadratic gravity without matter eq.(\ref{Rsq}) 
gives the  Einstein-Proca  action
 for the Weyl gauge field; this  became massive via Stueckelberg mechanism 
(spontaneous breaking). There is also a positive cosmological constant.
Conversely, the inverse gauge transformation of Einstein-Proca action takes one
to Weyl quadratic gravity action. Note again the conservation of the number 
of  degrees of freedom, impossible in the absence of Weyl gauge field.

To illustrate better  the Stueckelberg mechanism, 
write first eq.(\ref{Rsq2}) in  a Riemannian language
using  eq.(\ref{tildeR}) followed by  an integration by parts, which gives:
 \medskip
\bea\label{p}
L_1=\sqrt g\,
\Big\{-\frac{\xi_0}{2}\,\Big[\frac16 \phi_0^2\, R +(\partial_\mu\phi_0)^2\Big]
-\frac{\xi_0}{4!}\phi_0^4
+\frac{1}{8}\,\xi_0\,\phi_0^2\,\big(\w_\mu- \partial_\mu\ln\phi_0^2\big)^2
-\frac{1}{4 q^2}\,F_{\mu\nu}^{2}\Big\}.
\eea

\medskip\noindent
where we used  that $\sqrt g \, D_\mu \w^\mu=\partial_\mu(\sqrt g \,\w^\mu)$.
Then using  gauge transformation (\ref{ct}) with $\Omega=\xi_0\,\phi_0^2/(6 M^2)$,
one finds again eq.(\ref{EP}). It is then obvious how the first term becomes
the Einstein term in (\ref{EP}) and how the term of coefficient 1/8
 gives the mass term for $\hat\w_\mu$
(Stueckelberg mechanism) in  (\ref{EP}). Note there is no negative kinetic term (ghost) 
in eq.(\ref{p}).

The mass of Weyl gauge boson is near the  Planck scale ($\sqrt{3/2}\, q\, M$)  for 
a coupling $q$ not too 
small, and comes from the $\tilde R^2$ term alone. Below this mass scale
this  field  decouples,  Weyl connection becomes Riemannian ($\w_\mu=0$) and Weyl 
quadratic action becomes  Einstein-Hilbert action. So Einstein gravity is 
just a ``low energy''  limit (broken phase)  of Weyl gravity. Then  previous, 
long-held criticisms of Weyl quadratic gravity are avoided; the effects mentioned earlier,
 associated with Weyl geometry, are  suppressed by a large value of the Weyl 
``photon'' mass $\propto\,$Planck scale.  
Then any change of the spacing  of the atomic spectral lines is suppressed 
by this high scale and can be safely ignored.

The result in (\ref{EP}) is in the Einstein gauge  of constant  $\phi_0^2=6 M^2/\xi_0$ 
which coincides with the Weyl gauge (of constant Weyl scalar curvature) 
since we saw  $\langle\phi_0\rangle^2=-\cR$, so on the ground state $\phi_0^2=(-\cR)=6 M^2/\xi_0$,
see also \cite{S2} for a discussion.
Actually, for a Friedmann-Robertson-Walker universe,  the scalar  field  naturally evolves  in time to 
$\phi_0=$constant   because of  a conserved current   $J_\mu=\phi_0\partial_\mu\phi_0$ \cite{Fe}. 
The  Planck scale thus  emerges naturally as  the  scale where  Weyl gauge symmetry is 
spontaneously broken.

\subsection{A more general case}

In a most general case, Weyl quadratic gravity can contain another independent
term\footnote{A Gauss-Bonnet (total derivative) term of Weyl geometry can also be present
\cite{T2},  not relevant here.}$^,$\footnote{The Weyl tensor squared term we included here
is usually required at the  quantum level.}
\bea\label{action}
L_1^\prime &=&\sqrt g\,
\Big\{\,\frac{1}{\eta}\,
\tilde C_{\mu\nu\rho\sigma} \,\tilde C^{mu\nu\rho\sigma}
+
\frac{\xi_0}{4!}\,\cR^2
\Big\}
\nonumber\\[-4pt]
&=&
\sqrt g\,\Big\{
\frac{1}{\eta}\,\Big[ C_{\mu\nu\rho\sigma}\, C^{\mu\nu\rho\sigma}+\frac32  \,F_{\mu\nu}^2\Big]
+\frac{\xi_0}{4!}\,\cR^2\Big\}
\eea 
where  $\tilde C_{\mu\nu\rho\sigma}$ and ($C_{\mu\nu\rho\sigma}$)  is the 
 Weyl tensor in Weyl geometry (Riemannian geometry), respectively;
these tensors are related  as shown above \cite{T2} with $F_{\mu\nu}$ 
the field strength of   Weyl gauge boson.
 Notice that in this case $F_{\mu\nu}^2$ term is automatically
present, so there is no need to add it ``by hand'' (on symmetry grounds as in (\ref{Rsq})); however, 
for canonical gauge kinetic term  one has in this case  $q^2=-\eta/6$ ($\eta<0$). 
The result of eq.(\ref{EP}) is still valid  since both Weyl tensors are invariant under 
Weyl gauge transformations; then the final Lagrangian contains an additional 
term  $C_{\mu\nu\rho\sigma}^2$; this is needed anyway 
at the quantum level when trying to renormalize 
SM in the presence of  gravity in (ungauged) local conformal models \cite{TH2}. 
In this  case and in the absence of a separate kinetic term for $\w_\mu$ in the first line of 
(\ref{action}) (allowed by the symmetry),
the mass of the Weyl gauge field $m_\w^2\sim q^2 M^2 \sim (-\eta)\, M^2$ is thus related
to the mass of the spin-two ghost contained in $C_{\mu\nu\rho\sigma}^2$. 
At this scale  the non-metricity of Weyl geometry steps in to modify the 
Levi-Civita connection.

\subsection{Adding matter}\label{matter}

Consider now Weyl quadratic gravity,  eq.(\ref{Rsq}), coupled  to a  matter scalar $\phi_1$:
\medskip
\be\label{ll1}
L_2=\sg\,\Big[\,
\frac{\xi_0}{4!}\,\cR^2 - \frac{1}{4 q^2}\,%g^{\mu\rho} g^{\nu\sigma}
F_{\mu\nu}^2 % F_{\rho\sigma}
\Big]
- \frac{\sg}{12} \,\xi_1\phi_1^2\,\cR
+
\sg\,\Big[\frac12 \,g^{\mu\nu}\,\tilde D_\mu\phi_1 \tilde D_\nu \phi_1 -
 \frac{\lambda_1}{4!}\phi_1^4\Big],
\ee

\medskip\noindent
which is invariant under (\ref{ct}) and the potential for $\phi_1$ 
is the only allowed by this  symmetry.

As  in eq.(\ref{Rsq2}),   replace  $\cR^2\!\ra\! -2\phi_0^2 \cR -\phi_0^4$, to obtain a classically
equivalent action
\medskip
\be\label{ll1prime}
L_2=\sg\,\Big[\,
- \frac{1}{2} \,\rho^2\,\cR
- \frac{1}{4 q^2}\,F_{\mu\nu}^2 % F_{\rho\sigma}
+
\frac12 \,g^{\mu\nu}\,\tilde D_\mu\phi_1 \tilde D_\nu \phi_1
-\cV(\phi_1,\rho)
\Big],
\ee
with
%\medskip
\bea\label{toE}
\cV(\phi_1,\rho)=\frac{1}{4!}\,\Big[\,
\frac{1}{\xi_0} \big(6\rho^2-\xi_1\phi_1^2\big)^2
+
\lambda_1\phi_1^4\,\Big],\qquad\text{with}\qquad
\rho^2=\frac16\,\big(\xi_1\phi_1^2+\xi_0\phi_0^2),
\eea

\medskip\noindent
where we replaced $\phi_0$ by  $\rho$. 
Using eq.(\ref{RR}), a Weyl gauge transformation~(\ref{ct})  with $\Omega=\rho^2/M^2$
 followed by (\ref{tildeR}) that introduces Riemannian 
$\hat R$, gives 
\medskip
\be\label{W3}
L_2=
\sqrt{\hat g}\, \Big\{
-\frac{1}{2} M^2\,\Big[\hat{R}-\frac{3}{2} \hat\w_\mu\hat\w^\mu\Big]
-\frac{1}{4 q^2} % \hat{g}^{\mu\rho} \hat{g}^{\nu\sigma} 
\hat F_{\mu\nu}^2 % \hat F_{\rho\sigma}
+\frac{\hat g^{\mu\nu}}{2}\hat{\tilde D}_\mu\hat\phi_1 \hat{\tilde D}_\nu \hat\phi_1
-\cV(\hat\phi_1,M)
\Big]\Big\},
\ee

\medskip\noindent
with ${\hat{\tilde D}}_\mu\hat\phi_1=(\partial_\mu-1/2\,\,\hat\w_\mu )\hat\phi_1$ and $M$ identified with
$M_{\text{Planck}}$.
As in the case without matter, we obtained the Einstein-Proca action of a  gauge field 
that became massive after  Stueckelberg 
mechanism  of ``absorbing'' the   dilaton $\ln\rho$. The mass of $\w_\mu$ is
 $m_\w^2=(3/2) q^2 M^2$ (after rescaling $\hat\w_\mu\ra q\,\hat\w_\mu$ in (\ref{W3})). 
A canonical kinetic term of $\phi_1$ remains, since  only one 
degree of freedom (radial direction $\rho$) is ``eaten'' by the
 vector field, see Section~\ref{WP}.

Under the same gauge transformation (``gauge fixing'') the initial potential $\phi_1^4$ becomes
\bea\label{fff}
\cV= \frac{3 M^4}{2\,\xi_0} 
\Big[1-\frac{\xi_1\hat\phi_1^2}{6 \,M^2}
\Big]^2
+
\frac{\lambda_1}{4!}\,\hat\phi_1^4.
\eea

\medskip\noindent
We have a negative  mass term ($m_{\hat \phi_1}^2=-\xi_1 M^2/\xi_0$) if  $\xi_1>0$. 
This originates in (\ref{toE}) due to the initial dilaton contribution to the 
potential $\propto\phi_0^4$ (coming from $\tilde R^2$),  with $\phi_0$ replaced   by 
$(6\rho^2-\xi_1\phi_1^2)$ and $\rho$  ``gauge fixed'' to $M$.
Then, if massless $\phi_1$ gives  a {\it positive}  contribution $\xi_1\phi_1^2$ 
to the  Planck scale ($M$) ($\xi_1>0$) this effect is ``compensated'' by a  
{\it negative}  contribution to its mass term in the  potential 
(and vice-versa) in (\ref{toE}). The original dilaton (in $\tilde R^2$)
plays a mediator role in bringing this negative contribution.
It is then interesting that both mass scales of the theory,  Planck scale 
and the scale $\langle\hat\phi_1\rangle$  are simultaneously generated by 
the same ``gauge fixing'' transformation~(\ref{ct}).

If $\phi_1$ is the higgs field, $\langle\phi_1\rangle$ is the  electroweak (EW) scale, 
then Stueckelberg mechanism also triggers EW breaking. This discussion 
remains valid for more matter fields $\phi_j$, in eqs.(\ref{ll1}) to (\ref{fff}) 
simply replace $\xi_1\phi_1^2 \ra \sum_j\xi_j\phi_j^2$ and 
$(\tilde D_\mu\phi_1)^2\ra \sum_j (\tilde D_\mu\phi_j)^2$.

\subsection{Canonical action, more scalar fields and SM in Weyl geometry}\label{more}

The Weyl-covariant derivative acting on $\hat\phi_1$ in (\ref{W3})  is a remnant of Weyl gauge 
symmetry, now broken. To have  a  ``standard''  kinetic term for $\phi_1$ i.e. remove 
couplings  $\hat\w^\mu \partial_\mu \hat\phi_1$ (similar to electroweak ``unitary 
gauge'') one can now do a field redefinition
\medskip
\bea\label{tt}
\hat\w_\mu^\prime =\hat w_\mu- \partial_\mu \ln \big(\hat \phi_1^2+6 M^2\big),\qquad
%\nonumber\\
\hat \phi_1= M\sqrt{6}\,\sinh\Big[\frac{\h}{M\sqrt 6}\Big]
\eea
to find\footnote{
In terms of the initial fields $\w_\mu$ and $\phi_{0,1}$ eq.(\ref{tt}) can  be written as:
$\hat\w_\mu^\prime=\w_\mu- \partial_\mu\ln K$ where we denoted $K\equiv \xi_0\phi_0^2+(1+\xi_1)\phi_1^2$ 
and there is a current $J^\mu=(-1/4)\, g^{\mu\nu} \nabla_\nu K=(-1/4)\, g^{\mu\nu}
(\partial_\nu - \w_\nu)\,K$, which is a total derivative and   is conserved $\nabla_\mu J^\mu=0$. 
This is shown by applying $\partial_\mu$ on the equation of motion for
$\w_\mu$: $q \sqrt g J^\mu+\partial_\rho (\sqrt g F^{\rho\mu})=0$. 
Notice that one can also write the current as $J^\mu=(1/4)\,K\,\hat\w_\mu^\prime$.}
\be\label{twp}
L_2=\sqrt{\hat g}\, \Big\{
-\frac{1}{2} M^2\,\hat{R}
+ \frac34  M^2 \cosh^2 \Big[\frac{\h}{M\sqrt 6}\Big] \,\hat\w^\prime_\mu\hat\w^{\prime\,\mu}
-\frac{1}{4 q^2} % \hat{g}^{\mu\rho} \hat{g}^{\nu\sigma} 
\hat F^{\prime\, 2}_{\mu\nu}  % \hat F^\prime_{\rho\sigma}
+\frac{\hat g^{\mu\nu}}{2}\partial_\mu\h \partial_\nu \h
-{\hat \cV}\Big\}
\ee
with
\bea\label{scalars2}
{\hat \cV}&=&\frac{3}{2} \frac{M^4}{\xi_0} \Big[
1-\xi_1\sinh^2 \frac{\h}{M\sqrt 6}\Big]^2
+
\frac{3}{2} M^4 \,\lambda_1  \sinh^4\frac{\h}{M\sqrt 6}.\label{ov1}
\eea

\medskip\noindent
In (\ref{twp}) one finally rescales $\hat\w^\prime_\mu\ra q\,\hat\w_\mu^\prime$ 
for a canonical gauge kinetic term.

Taylor expanding the mass term of the Weyl gauge field
 for small $\h<M$ shows there are  additional corrections to this mass
beyond those due to Stueckelberg mechanism, since $\langle\h\rangle\not=0$.
 Note there is no restriction in the action regarding the relative values
 of $\h$ versus $M$.  For $\h>M$, $\hat \cV$ is always positive if 
$\xi_1^2/\xi_0+\lambda_1>0$ which can be true even for  $\lambda_1<0$. 

This potential is relevant for models of inflation, assuming $\h$ is the inflaton. 
Since Planck  scale emerged as the scale where Weyl gauge symmetry is spontaneously broken, eq.(\ref{W3}), 
field values above this scale are natural, which is relevant for inflation.
For $\lambda_1$ and  $\xi_1$ very small, e.g. $\lambda_1 \xi_1\sim 5\times 10^{-10}$
 and $\xi_1\sim 10^{-5}-10^{-3}$,  the potential is nearly  flat and one has inflation 
similar to  Starobinsky model \cite{Sta} for suitable $\xi_0\approx 10^{10}$, see  \cite{Barnaveli}
 for the analysis of this potential. The larger quoted values of $\xi_1$ mark a departure
from the Starobinsky inflation.  But unlike in \cite{Barnaveli} where no Stueckelberg mechanism
takes place since there is no gauge kinetic term,
 here there is no flat direction  left - this was ``absorbed'' by the gauge field
which became massive. This issue and inflation 
are discussed in detail elsewhere \cite{Winflation}.

The situation here is  also different from the common models of inflation  of
no matter field  present with inflation driven by $\sqrt{g}\, (R^2+M^2 R)$.
Here the ``scalaron'' mode is actually a compensator eaten by 
$\w_\mu$, while the matter field $\h$ is the inflaton. 
Further, if there is no Weyl gauge field in (\ref{ll1}), (set $\w_\mu=0$), 
 inflation is still possible and was already studied in \cite{Rinaldi}.
The scenario  is again  similar to that in Starobinsky models. Finally, in the
absence of the quadratic term, with only a linear term in scalar curvature and  
global scale invariance, inflation is again possible and was discussed in  \cite{Fe,GGR}.

The analysis so far can be extended to more scalar fields present in a Weyl gauge invariant action.
 For example, for
 two scalar fields in eq.(\ref{ll1}) with non-minimal couplings $\xi_1$, $\xi_2$, 
and with an initial potential $V(\phi_1,\phi_2)$ replacing that in (\ref{ll1}),
one obtains  the following canonical action, similar to that in (\ref{twp}):
\medskip
\be
L_2=\sqrt {\hat g}\,\Big\{
\frac{-1}{2}\,M^2\,\hat R -\frac{1}{4 q^2}\,\hat F^\prime_{\mu\nu}\hat F^{\prime \,\mu\nu}\!
+\frac12 m^2(\sigma) \,\hat\omega_\mu^\prime \hat\omega^{\prime \mu} 
\!+\frac12\Big[\sinh^2\!\!\frac{\sigma}{M\sqrt 6}\,(\partial_\mu\tilde\theta)^2
\!+\! (\partial_\mu\sigma)^2\Big] -\hat \cV\Big\}
\ee

\medskip\noindent
with the notation $m^2(\h)=(3/4) M^2 \cosh^2 (\h/(M\sqrt 6))$ and
with new polar coordinates fields, 
$\tan\theta=\phi_1/\phi_2$ and $\phi_1^2+\phi_2^2=6 M^2\sinh^2 
\sigma/(M\sqrt{6})$ and
$\tilde \theta=M\sqrt 6\, \theta$. Finally, the potential is
%\medskip
\be\label{ov2}
\hat \cV= \frac{3\,M^4}{2\, \xi_0}
\Big[1- \xi_{12}\,\sinh^2 \frac{\h}{M\sqrt 6}\Big]^2
+
\frac{3}{2} M^4 \big(24\,V(s_\theta,c_\theta)\big)\sinh^4 \frac{\h}{M\sqrt 6},
\ee
%
%\medskip\noindent
where $\xi_{12}=(\xi_1\sin^2\theta+\xi_2\cos^2\theta)$ and
 $s_\theta\!=\!\sin\theta$, $c_\theta\!=\!\cos\theta$.
For an  $\cO(2)$ symmetry, $\theta$ dependence in the potential disappears so one can introduce
$V(s_\theta,c_\theta)\!=\!\lambda/4!$ but kinetic mixing remains. Compare (\ref{ov1}), (\ref{ov2})
to notice the similar structure of the potential.

These results make it attractive to consider the  Weyl gauge symmetry 
for model building beyond SM and Einstein gravity. 
With Einstein gravity as  a ``low energy'' broken phase  of Weyl quadratic gravity, then
Weyl  gauge symmetry and Weyl gravity are  ``freed''  from past criticisms based on the 
(wrong) assumption   that the  Weyl gauge field is  massless.
One can consider the SM with a higgs mass parameter set to zero and
extend  it with Weyl gauge symmetry. In such case note that,  interestingly,
only the SM Higgs/scalars couple to  the Weyl gauge boson, as in 
(\ref{ll1}) or equivalently (\ref{twp}).  The
 SM fermions do not couple to $\w_\mu$ \cite{Kugo,M,ghilen}.
The SM gauge fields kinetic terms are also invariant under Weyl gauge symmetry 
see e.g. \cite{ghilen}.
Therefore, the SM Lagrangian formulated in  Weyl conformal geometry is
\bea
L=L_2+ L_{f+g}^{SM}
\eea
with $L_2$ as in (\ref{ll1}) adapted for the Higgs sector and 
other scalar fields (e.g. inflaton), as above. $L_{f+g}^{SM}$ denotes the usual SM Lagrangian
for the fermionic and gauge sectors.
Note however that a  kinetic mixing
of $\w_\mu$ with $U(1)_Y$ of SM  ($\sqrt{-g}\, F^\w_{\mu\nu} F^{\mu\nu}_Y$)
 is allowed by the SM and Weyl gauge symmetries.
 This has implications for phenomenology not yet explored\footnote{
This mixing can be neglected for a large enough non-metricity scale (mass $m_\w$).}.

\subsection{Related models}

There is a difference between the Weyl gauge invariant model discussed here
 and the case  of local conformal extensions (no gauging)
 of the Standard Model, see e.g.\cite{TH,TH2,Bars},
 when generating the Planck scale spontaneously (by dilaton vev).
As we saw, a Weyl gauge invariant model conserves the number of degrees of freedom $n_{df}$
during the  breaking of this symmetry. 
Moreover, there is  no ghost field in Section~\ref{QG} when ``gauge fixing''
 the Planck scale in eqs.(\ref{EP}), (\ref{p}).
 This is to be compared to these local conformal extensions of the 
SM where the Einstein term $(-\sqrt g/2) M^2 R$  is written in a local conformal invariant
way as
%\medskip
\bea
\cL_E=\sqrt g\,\Big\{
- \frac{\xi_0}{2} \,\Big[\,\frac{1}{6}\,\phi_0^2\,R+(\partial_\mu\phi_0)^2\Big]
\Big\},
\eea

\medskip\noindent
to be compared to (\ref{p}).
We see here that trying to generate  Planck scale as a vev of $\phi_0$, by ``gauge fixing''
$\phi_0^2=6 M^2/\xi_0$,  demands the notorious negative kinetic term (ghost dilaton) be present
(see \cite{Oh,ghilen} for  a discussion).
Also, in this local conformal case, 
when ``gauge fixing''  $\phi_0$ to a constant  (``unitary gauge'') and this symmetry is broken,
there is no gauge field to ``absorb'' this scalar (dilaton) mode, see e.g. \cite{TH2}.
Therefore $n_{df}$ is not conserved and shows the need for the Weyl gauge symmetry,
for self-consistency.

Models with Weyl gauge symmetry seem to be allowed by  black-hole physics.
This is not true for models with global  symmetries, in particular  
global scale symmetry (e.g. A-gravity \cite{Ag}) since
global charges can be eaten by black holes which subsequently evaporate \cite{KA}.
Further, in models with Weyl gauge symmetry 
higher dimensional/curvature operators (beyond quadratic ones of dimension $d=4$ in the Weyl action)
cannot be present since they should be suppressed by some high scale not present 
in the theory (forbidden by this symmetry).
  Also, the dilaton  is eaten by the Weyl ``photon''
which becomes massive, so dilaton powers cannot be present to suppress 
such effective operators either. This may remain true at the quantum level, assuming quantum calculations
respect this symmetry. 
This requires an ultraviolet   regularization that preserves the Weyl gauge symmetry
\cite{En}. %,Sh}.

This analysis  can be extended to other  non-metric theories, with torsion, etc.
Our result is in agreement with more general approaches  \cite{RP1} that consider that at 
 a fundamental level gravity is a  theory of connections as dynamical objects.
Some of these connections become massive (via Stueckelberg mechanism), as we saw  for 
the Weyl connection,  while Levi-Civita connection remains massless. 
In our case Weyl connection departed from the (fixed) Levi-Civita by a correction
$\w_\mu$ which was a dynamical field.
More generally, one can write any  dynamical connection as a  Levi-Civita
connection plus a tensor field contribution   which is a sum of
a non-metricity tensor  (here due to $\w_\mu$) and a contorsion tensor, and then re-do
this analysis. From a particle physics perspective, this tensor field, being massive,
should decouple and leave in the ``low energy'' limit only the Levi-Civita connection.

For high scale physics and  early cosmology,  non-metricity effects cannot
be ignored. In fact current lower bounds on non-metricity scale, which is set by the mass
of the Weyl field, are very low, in the region of few TeV \cite{La}. With 
the mass of Weyl field $\w_\mu$ of $\sqrt{3/2} \, q\, M_\text{Planck}$,
this region would correspond to ultraweak values of the coupling $q$. One may also explore
the possibility that $\w_\mu$ is a dark matter candidate.
In Weyl invariant  models  of vector dark matter (DM), the mass of the 
DM vector field  is again in the region of few TeV \cite{Ta} (also \cite{Ya}). 
This is interesting for phenomenology and deserves careful study.

The aforementioned  separation of the connection into metric and non-metric contributions
is also  useful for studies of asymptotic safety. These are using  the metric formalism 
(with Levi-Civita connection), so they do not take into account non-metricity effects, etc. 
Their results  could  however be extended by simply taking into account the 
new degrees of freedom (fields)  which are corrections to the Levi-Civita connection. 
Then  asymptotic safety in a non-metric theory is that of  a theory with Levi-Civita 
connection plus the dynamical effects of these fields.

\section{Conclusions}

In this work we studied the effect of  Weyl gauge symmetry  
beyond SM and Einstein gravity, in the context of Weyl conformal geometry.  
This geometry is of interest since it may play a role 
in early cosmology or at  high scales when effective field theory  becomes nearly conformal.   
To  take advantage of its  symmetry we used  
the  action expressed in   terms of tensors and scalar curvatures of Weyl geometry
(instead of their  Riemannian expressions) since these are Weyl-invariant 
 and covariant, respectively. 
 In this (Weyl) formulation, individual operators in the action are invariant 
under Weyl gauge symmetry.  Then  this symmetry and internal  gauge symmetries 
are on an equal footing in the action.

 Weyl conformal geometry  has a built-in geometric Stueckelberg mass mechanism.
By using this Weyl formulation we showed that: 
{\bf a)}  a simple Weyl  ``gauge fixing'' symmetry transformation easily transforms
an action in  Weyl geometry  directly 
into an  action in Riemannian geometry,
due to Stueckelberg breaking of the Weyl gauge symmetry; {\bf  b)}
in this step no fields re-definitions are used, only gauge transformations;  {\bf c)}
 no negative kinetic term (ghost) is generated
and the number of degrees of freedom is conserved
{\bf d)} Planck scale is an emergent scale where Weyl gauge symmetry 
is spontaneously broken; hence  field values above the Planck scale are natural;
{\bf e)} calculations simplify dramatically compared to a Riemannian formulation
of this symmetry.

To detail, there  is a  conservation of the number  of dynamical degrees of  freedom ($n_{df}=3$)
in step a) above, as required for spontaneous breaking: the initial massless  gauge field $\w_\mu$
(defining the Weyl connection) absorbs  the dilaton (compensator) and  becomes massive,
then decouples, hence  Weyl connection  becomes Levi-Civita connection.
Thus mass generation has
 a geometric interpretation as a transition from Weyl geometry to the Riemannian one.
Note that in the (ungauged) local conformal models, a similar ``gauge fixing'' 
(of the dilaton to a constant vev), 
does not conserve $n_{df}$  when the  symmetry breaking takes place, 
since there is no vector field to ``absorb'' the Goldstone mode of the symmetry. 

Using this idea for the original Weyl quadratic gravity,
 one finds that this  action is  
immediately transformed  by a ``gauge  fixing''  symmetry transformation,
into  Einstein-Proca  action for the  Weyl gauge field 
plus a (positive) cosmological constant and matter action (if initially present); 
the Weyl gauge field undergoes   a Stueckelberg mechanism.
Below its  mass ($\sim q M_\text{Planck}$) this field decouples,
hence Einstein gravity is simply a ``low energy'' {\it broken phase} of Weyl quadratic gravity.
No ghost field is present, in contrast with the (ungauged) local conformal models.

Past  criticisms of Weyl gravity, related to non-metricity,
assumed the Weyl gauge field to be massless; 
these criticisms are   avoided since such effects induced by the Weyl gauge 
field  are actually strongly suppressed by its  mass
expected to be high (for $q$ not too small). However, note that current lower
 bounds on the non-metricity scale
 ($m_\w$) are   low (TeV region). This suggests  that 
the Weyl field could  in principle be lighter, if one considers ultraweak values of 
the coupling~$q$, and even act as a dark matter candidate. This would be a ``geometric'' solution
to dark matter since $\w_\mu$ is part of the original Weyl geometry.
This is interesting and deserves careful study.

When building Lagrangians with Weyl gauge symmetry, only scalar
fields (e.g. Higgs sector of the SM) couple to the Weyl field $\w_\mu$.
Following the same ``gauge fixing'' transformation, there exists a
``compensating'' mechanism for matter scalars  with non-minimal couplings 
to $\tilde R$: if a massless scalar  gives  a {\it positive} (negative)
 contribution to the generation of the Planck scale, this is
``compensated'' by   a simultaneous {\it negative} (positive) mass squared 
term, i.e. a spontaneous breaking of the symmetry
under which it is charged.  This is due to a  dilaton term in the potential 
induced when ``linearising'' the quadratic Weyl scalar curvature term.

Models with Weyl gauge symmetry seem to be  allowed by
black-hole physics, unlike models with {\it global} 
scale symmetry (e.g. Agravity).  Further, in models with Weyl gauge symmetry 
higher dimensional/curvature operators, beyond the quadratic ones of $d=4$ of the Weyl action, 
are forbidden since they should be suppressed by some high scale not present 
in the theory and forbidden by this symmetry.
  Also, the dilaton (compensator) is eaten by the Weyl ``photon''
which becomes massive, so such effective operators could not be suppressed by powers
of the dilaton either.
This may remain true at the quantum level, assuming quantum calculations
respect this symmetry. This is relevant for attempts to prove renormalizability
of Weyl gravity action.

Our results may also be of interest to asymptotic safety theories; these
are using the metric formalism (Levi-Civita connection)
and miss the effects discussed in this work. However, these studies
can be extended to apply here by taking into account the dynamics of the 
new fields ($\w_\mu$) that are corrections to the Levi-Civita connection. 
So  asymptotic safety in non-metric case is that for Levi-Civita connection
plus the additional fields dynamics. 

These results indicate that the original Weyl quadratic gravity 
is physically relevant and its role  should be  reconsidered,
together with its implications for other  areas:  
SM extended  with Weyl gauge symmetry, its supersymmetric version, 
black-hole physics and cosmology\footnote{
As stated  by Weyl long ago \cite{Weyl2}:
``The action [...] that was implemented in the previous sections is constituted as [...]
a linear combination of $\cR^2$ and $F_{\mu\nu}^2$. I believe that
one can assert that this action principle implies everything that Einstein’s theory has
implied up to now, but in the more far-reaching questions of cosmology and the
constitution of matter, it exhibits a clear superiority.  Nevertheless, I do not believe that
the laws of nature that are exactly applicable in reality are resolved by it. ''}.

\end{document}